# Very Asymmetric Collider for Dark Matter Search below 1 GeV

B. Wojtsekhowski, V.S. Morozov, and Y.S. Derbenev

*Jefferson Lab, Newport News, VA 23606*

As illustrated in Fig. 1 [1], current searches for a dark photon in the mass range below 1 GeV require an electron-positron collider with luminosity of the level of at least $10^{34}$ cm$^{-2}$s$^{-1}$. The challenge is that, at such low energies, the collider luminosity rapidly drops off due to increase in the beam sizes, strong mutual focusing of the colliding beams, and enhancement of collective effects. However, recent advances in accelerator technology including the nano-beam scheme of SuperKEK-B [2], high-current Energy Recovery Linacs (ERL) [3], and magnetized beams [4] allow one to achieve a luminosity $L$ of $> 10^{34}$ cm$^{-2}$s$^{-1}$ at the center of momentum energy of $< 1$ GeV.

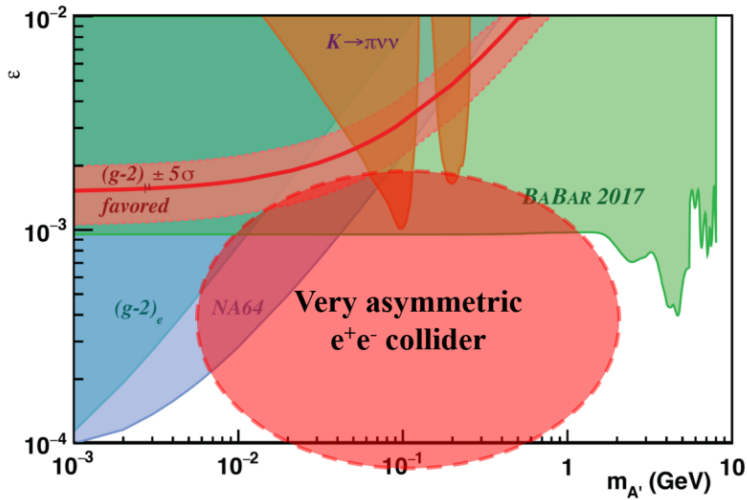

Fig. 1: Mixing constant vs. heavy photon mass [1].

We propose such a facility based on a positron storage ring and an electron ERL. Such a facility is illustrated in Fig. 2 using the positron storage ring of SuperKEK-B and the Cornell ERL project as examples. Note that the scale of the ERL is increased by a factor of 10 or so for better visibility. The positron storage ring of a few GeV provides a high-current, high-quality positron beam. The storage ring configuration allows for accumulation of a few Amps of positron beam while its relatively high energy ensures damping of the beam to small emittances $\varepsilon$ and makes the beam less sensitive to focusing by the electron beam. On the other hand, an electron beam from an ERL has low energy, modestly high current, high beam quality and can be perturbed much stronger by the oncoming beam than a stored beam. An ERL rather than a straight-through linac is needed for high energy efficiency at a high electron current. A combination of these properties allows for a high luminosity at a low center of momentum energy.

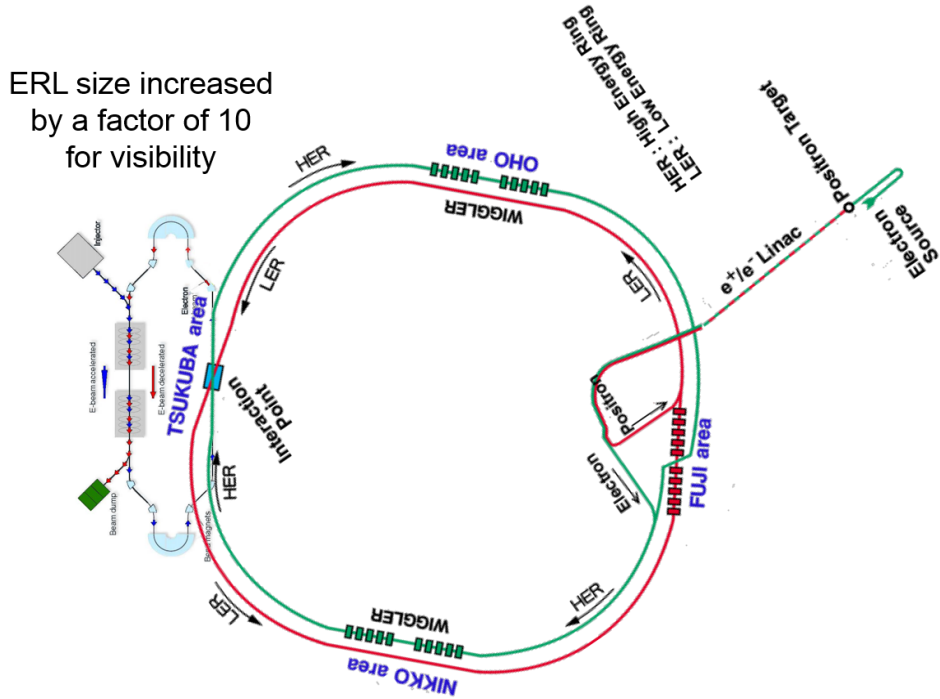

Fig. 2: Layout of the proposed ERL-ring facility based on the SuperKEK-B [2] positron storage ring and Cornell ERL project [3].

The fundamental limitations on the luminosity come from the maximum achievable beam-beam tune shift $\xi$ of the stored beam, which is a measure of focusing provided by the opposite beam, and from an analogous quantity for a linac beam called the disruption parameter $D$. To reach a high luminosity while keeping these parameters within demonstrated limits, we proposed to use the SuperKEK-B nano-beam scheme. The beams are over-focused and are colliding at an angle $\theta$. This provides high luminosity from the overlapping beam waists while keeping the mutual beam perturbations low.

The nano-beam scheme works best for flat beam while ERL beams tend to be round. Therefore, we propose to use an ERL with a magnetized beam [4]. A magnetized beam is produced inside the magnetic field of a solenoid. It is round but can have a ratio of the canonical emittances of up to 100. After acceleration, such a beam is converted to a flat beam with the canonical emittances transforming to the usual horizontal and vertical emittances as illustrated in Fig. 3 [5]. We assume an emittance ratio of a factor of 20. The parameters of the proposed linac-ring facility are summarized in Table 1. The facility parameters are compared to those of SuperKEK-B and the Cornell ERL project. As one can see, with all parameters being within demonstrated limits, the achievable luminosity is $1.7 \times 10^{34}$ cm$^{-2}$s$^{-1}$, which is quite suitable for a dark photon search in the mass region of $< 1$ GeV.

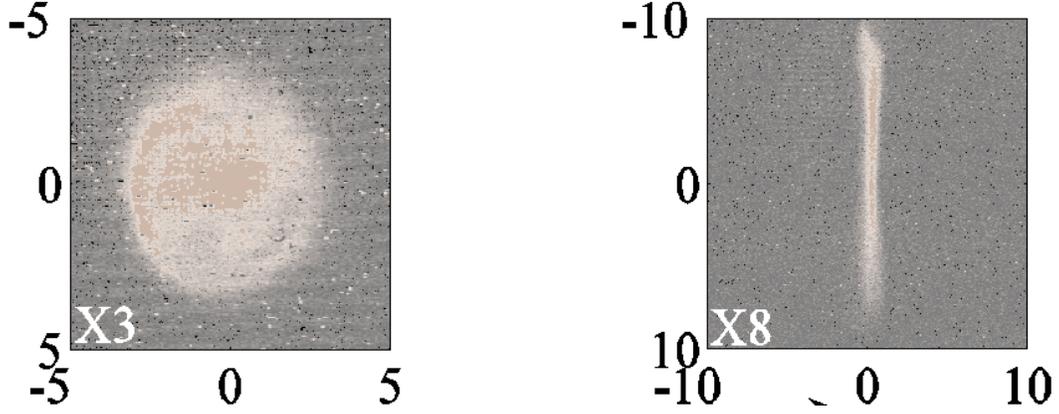

Fig. 3: Measured magnetized beam before (left) and after (right) round-to-flat transformation [5].

Table 1: Parameters of the proposed ERL-ring facility and comparison to similar projects.

|  |  | SuperKEK-B LER | Cornell ERL | Linac-Ring | |
|---|---|---|---|---|---|
|  |  | $e^+$ | $e^-$ | $e^+$ | $e^-$ |
| $E$ | GeV | 4 | 5 | 4 | 0.1 |
| $f$ | MHz | 248.5 | 1300 | 248.5 | |
| $I$ | A | 3.6 | 0.1 | 3.6 | 0.095 |
| $\sigma_z$ | mm | 6 | 0.6 | 6 | 1 |
| $\varepsilon_x/\varepsilon_y$ | pm/pm | 3200 / 8.64 | 31 / 31 | 850 / 85 | 5100 / 255 |
| $\varepsilon_x^N/\varepsilon_y^N$ | μm/μm | 25 / 0.07 | 0.3 / 0.3 | 6.7 / 0.67 | 1 / 0.05 |
| $\beta_x^*/\beta_y^*$ | mm | 32 / 0.27 |  | 6 / 0.6 | 1 / 0.2 |
| $\sigma_x^*/\sigma_y^*$ | μm/μm | 10 / 0.048 |  | 2.3 / 0.23 | 2.3 / 0.23 |
| $\theta$ | mrad | 83 |  | 10 | |
| $\xi_x/\xi_y$ |  | 0.0028 / 0.088 |  | 0.016 / 0.016 | |
| $D_x/D_y$ |  |  |  |  | 1.7 / 17 |
| $R_{h.-g.}$ |  |  |  | 0.9 | |
| $R_{c.-a.}$ |  |  |  | 0.7 | |
| $L$ | cm$^{-2}$s$^{-1}$ | $8 \times 10^{35}$ |  | **Achievable $1.7 \times 10^{34}$** | |

The energies below 1 GeV can be covered by lowering the linac energy. Keeping the luminosity at the same level requires increase in the linac current inversely proportional to energy. The total electron beam power remains fixed and none of the limiting collision parameters are exceeded.

Realization of the proposed project could start with an R&D stage at any facility with an available positron storage ring, e.g. at Frascati or Cornell.

## Acknowledgements

Authored by Jefferson Science Associates, LLC under U.S. DOE Contract No. DE-AC05-06OR23177. The U.S. Government retains a non-exclusive, paid-up, irrevocable, world-wide license to publish or reproduce this manuscript for U.S. Government purposes.

## References

[1]   J. P. Lees *et al.*, arXiv:1702.03327 [hep-ex] (2017).
[2]   Y. Ohnishi *et al.*, Prog. Theor. Exp. Phys. (2013) 2013 (3): 03A011.
[3]   C.E. Mayes *et al.*, in Proc. PAC'11, TUOBS2, p. 729 (2011).
[4]   A. Burov *et al.*, Phys. Rev. E 66, 016503 (2002).
[5]   Y.-E Sun *et al.*, Phys. Rev. ST Accel. Beams 7, 123501 (2004).